\documentstyle[12pt]{article}
\newcommand{\tr}[1]{\,{\rm tr}\,#1\,}
\newcommand{\p}[1]{\partial}
\begin{document}
\title{\begin{flushright}
{\small SMI-15-92 \\ December, 1992 }
\end{flushright}
\vspace{2cm}
Large-N Quenching in the Kazakov-Migdal Model.}
\author{I.Ya. Aref'eva \thanks{E-MAIL: Arefeva@qft.mian.su}
\\ Steklov Mathematical Institute,\\ Russian Academy of Sciences,\\
\\ Vavilov st.42, GSP-1,117966,\\ Moscow, Russia }
\date{~}\maketitle
\begin {abstract}
To study the behavior of the  Kazakov-Migdal at large N the  quenched momentum
prescription with constraints for treating the large N limit of gauge theories
is used. It is noted that it leads to a quartic dependence of an action on
unitary  matrix instead of a quadratic dependence discussed in previous
considerations. Therefore
the model is not exactly solvable in the weak coupling limit.
An approximation procedure for investigation of the model is outlined. In this
approximation  an indication to a phase transition
for $d<4,8$ with $\beta _{cr}=\frac{1}{d-4,8}$ is obtained.
\end {abstract}
\newpage
\section{Introduction}
Recently Kazakov  and Migdal \cite {KM}  have made an important progress in an
attacking  the old problem of an evaluation of the large N limit for gauge
models
\cite {W80}. They proposed a lattice
gauge model which is solvable in the large  N limit under an assumption of
translation
invariance of a master field and has a nontrivial critical behaviour. Different
aspects of the Kazakov  and Migdal (KM) model were considered \cite {Mig92a} -
 \cite {D'AT}. A relation of this model to QCD and, in particular, the property
of asymptotic freedom still should be clarified.

In this note a standard quenched momentum  prescription  \cite {EK} -
\cite {Das} for treating the large N  limit is used for the KM model .
It has been established  \cite {BHN,GK} that in the case of gauge theories the
quenching of the momentum must be accompanied by a constraint on the
eigenvalues
of the covariant derivative. It will be shown that this constraint leads to a
quartic dependence of an action on unitary matrix
instead of a  quadratic dependence discussed in previous considerations. In the
strong coupling limit one can expect that the constraint can be omit and one
lefts with an action which is quadratic on the unitary matrix $V$.
In this case we can identify the translation
invariant master field with a reduced field without quenching, i.e. without
constraints. The relation of the translation  invariant master field with the
reduced field without quenching was noted  by Makeenko \cite{Ma}.  However one
cannot ignore the constraint in all regions of the coupling constant. This
drops
a hint that the approximation of the constant master field does not fully
describes an asymptotic of the KM model at the large N limit. This constraint
makes the KM model  not exactly solvable for the weak coupling. We calculate
the
integral over unitary matrix in the semiclassical approximation, i.e.
at small coupling. By the analogy with the  Gross-Witten  \cite {GW} and the
Brezin-Gross  \cite {BG} models it is natural to expect a phase transition for
a theory with an action being a polynomial on the one link unitary , in
particular,
for the quartic action. In the framework of some approximation procedure
scheme this phase transition gives rise to a phase transition for the KM model.

The paper is organized as follows. In Section 2 the  general quenched momentum
prescription with constraints  is applied to the KM model and a reduced
action with quartic dependence on the  gauge field is presented.
Then, in Section 3 we integrate out the gauge fields in the semiclassical
approximation. In Section 4
we use an approximation scheme to evaluate the remaining integrals over the
quenched momentum and the eigenvalues of the scalar field.
The concluding remarks are collected in the last Section.

\section   {Quenched Momentum Prescription for the KM model}
The KM  lattice gauge model is defined by
the partition function
\begin {equation} 
                                                          \label {1.1}
Z_{KM}=\int \prod_{x,\mu} dU_{\mu}(x) \prod_x d\Phi(x)
e^{\sum_x N \tr{\left(-V[\Phi(x)]+
\beta\sum_{\mu}\Phi(x)U_\mu(x)\Phi(x+\mu)U_\mu^\dagger(x)\right)}}.
\end   {equation} 
Here the field $\Phi(x)$ takes values in the adjoint representation of the
gauge group $SU(N)$ and the link variable $U_\mu(x)$ is an element of the
group, $\mu =1,...d$.

The well known procedure for treating the  large N limit of $N\times N$ matrix
models consists in
application of a quenched momentum prescription.
This prescription for treating  the limit of infinite N
 of theories with a global,
or local gauge, $U(N)$ symmetry, both on the lattice and in the continuum,
has been obtained more than
ten years ego by Eguchi and Kawai \cite {EK}, Bhanot, Heller and Neuberger
 \cite {BHN},  Parisi  \cite {Pa} and
Gross and Kitazawa  \cite {GK} (see also  \cite {GAO}- \cite {EN}).
Generally speaking, according to this prescription
a reduced model can be described as containing just one side (one space-time
point)
and one  scalar field $\Phi$ on this side, or d links attached to one side and
$d$  fields  for vector field. To get the reduced action one should replace
a matrix field $\Phi (x)$ with  $D(x)\Phi D^\dagger(x)$, where $D(x)=
e^{ip_{\mu}x_{\mu}}$, and
$p_{\mu}$ is the diagonal matrix with matrix elements $p_{\mu}^{i}, i=1,...N$.
Then, one gets the vacuum energy in the large N limit by integration the free
energy of obtained action over $p$. It has been shown that
quenched theory produces the standard Feynman diagrams for invariant
Green functions in all orders in perturbation theory.

In the case of gauge theories  the quenching of the momentum must be
accompanied by a constraint on the eigenvalues of the covariant derivative.
Without this constraint one gets naive reduced model  without quenching. In the
case of the Wilson gauge theory the naive reduced model describes correctly
the theory at large N only in the strong coupling regime.

Let us apply the quenched momentum prescription
to the KM gauge theory (\ref {1.1}). If we were to
follow the quenched momentum prescription we would replace $\Phi(x+\mu)$
with $D_{\mu}\Phi(x)D_{\mu}^\dagger$, where $D_{\mu}=e^{ip_{\mu}a}$, and
$p_{\mu}$ is the standard diagonal matrix whose elements are $|p_{\mu}^{i}|<\pi
/a$, $a$ is the lattice spacing.  The quenched model will then have an action
\begin {equation} 
                                                          \label {2.1}
\tilde{S}[\Phi , U_\mu ]=
(a)^{d}[-\tr V(\Phi)+\frac{1}{g^{2}}\sum_{\mu >0}
\tr\Phi U_\mu D_{\mu}\Phi D_{\mu}^\dagger U_ \mu^\dagger ],
\end   {equation} 
and the vacuum energy
\begin {equation} 
                                                          \label {2.2}
\tilde{F}(p) =\ln \left[\int d\Phi d \mu(U_{\mu}) \exp (-\tilde {S}(\Psi ,
U_{\mu},p_{\mu}))\right].
\end   {equation} 
The vacuum energy per unit volume at the large $N$ limit is then obtained by
integrating
$\tilde{F}(p_{\mu})$ over all values of $p_{\mu}$
\begin {equation} 
                                                          \label {2.3}
F=\lim_{N\to \infty}\int dp_{\mu} \tilde{F}(p_{\mu}).
\end   {equation} 
The integration over $p_{\mu}$ is normalized to unity for all values of N
\begin {equation} 
                                                          \label {2.4}
\int dp=\prod _{\mu =1}^{d}\int _{-\pi/2a}^{\pi/2a}\prod
_{i=1}^{N}[\frac{d^{d}p_{i}}
{(\pi/a)^{d}}] (\frac{1}{2a})^{d}.
\end   {equation} 
If we were used the standard Haar measure, $dU_{\mu}$,
then by a change of variables $U_{\mu}\to U_{\mu}D_{\mu}$ we would eliminate
$D_{\mu}$ from the action (\ref {2.1}). Thus no quenching would have occurred,
and we would
recover the KM model in the translationally invariant master field
approximation,
\begin {equation} 
                                                          \label {2.5}
F=\ln\left[\int d\Phi dU_{\mu} \exp (a^{d}N[-\tr V(\Phi)
+\beta\sum_{\mu >0}
\tr\Phi U_\mu \Phi  U_ \mu^\dagger ])\right].
\end   {equation} 
The same procedure being applied to the standard  Wilson theory yields
to the Eguchi-Kawai reduced model without quenching, which is known to be
necessary to obtain the correct results in weak coupling.
The exit from this problem is known, one has to introduce a gauge invariant
constraint on the $U_{\mu}$'s  \cite {GK}. The suitable constraint restricts
the eigenvalues of $U_{\mu}D_{\mu}$ to be equal to $D_{\mu}$. Note that in
continuum case the analog of these constraints yield the correct coupling
constant renormalization  \cite {GK}. We are going to use these constraints for
the reduced model (\ref {2.1}). This means that we have to take the measure
$d\mu (U)$ in (\ref {2.2}) to be
\begin {equation} 
                                                          \label {2.6}
d\mu (U)=\prod dU_{\mu} C(U_{\mu},D_{\mu}),
\end   {equation} 
where $dU_{\mu}$ is the Haar measure on $SU(N)$  and
\begin {equation} 
                                                          \label {2.7}
C(U_{\mu},D_{\mu}) =\prod _{\mu}\int dV_{\mu}~\Delta(D_{\mu})~
\delta(U_{\mu}-V_{\mu}D_{\mu}V_{\mu}^\dagger D_{\mu}^\dagger )
\end   {equation} 
$$\Delta(D_{\mu})=\prod _{i<j}\sin ^{2}(\frac{p_{\mu}^{i}-p_{\mu}^{j}}{2}a).$$
For the moment we omit the Faddeev-Popov determinant which should
be taken into account if one wants to get a correct result in weak coupling.
If we integrate out the $U_{\mu}$, setting
$U_{\mu}=V_{\mu}D_{\mu}V_{\mu}^\dagger D_{\mu}^\dagger $
we obtain
\begin {equation} 
                                                          \label {2.8}
\exp(\tilde{F}(p)) =\int d\Phi \prod _{\mu} d V_{\mu}
\exp \{a^{d}N[-\tr V(\Phi)
+\beta\sum_{\mu >0}
\tr (\Phi V_{\mu}D_{\mu}
V_{\mu}^\dagger \Phi V_{\mu} D_{\mu}^\dagger V_{\mu}^\dagger )]\}.
\end   {equation} 
To find the vacuum energy at the large $N$ limit one should integrate
$\tilde{F}(p)$ over $dp_{\mu}$. In fact we will use the formula
\begin {equation} 
                                                          \label {2.8'}
F=\int d\mu (p)\tilde{F}(p),
\end   {equation} 
where
\begin {equation} 
                                                          \label {2.21}
d\mu (p)=\prod _{i}\frac{d^{d}p_{i}}{(\frac{\sqrt{\pi}}{a})^{d}}
\exp (-p^{2}_{i}a^{2}).
\end   {equation} 

In  comparison with equation (\ref {2.5}), where we have the integral from the
exponent
containing the quadratic dependence from the unitary matrix $U$, in (\ref
{2.8})
we have to integrate the exponent with the quartic dependence from the unitary
matrix.

\section   {Weak Coupling for the Reduced  KM Model with Quenching}
The action in (\ref {2.8}) contains the quartic  interaction between the
unitary matrix $V$ and the hermitian matrix $\Phi$, and the model does not look
the exactly solvable.
It is clear that the integral $I(\Phi ,D)$
 \begin {equation} 
                                                          \label {2.8''}
I(\Phi ,D) =\int  d V \exp [\beta \tr (\Phi V D
V ^\dagger \Phi V  D ^\dagger V ^\dagger ]
\end   {equation} 
depends only on eigenvalues of the hermitian
matrix $\Phi$, $\phi _{i}$ .

We are going to calculate the integral over $V$'s in (\ref {2.8}) in the
semiclassical approximation.
  To this end let us note that the corresponding
classical equation has a form
$$D V_{\mu}^\dagger \phi  V_{\mu} D^\dagger  V_{\mu}^\dagger \phi -
V_{\mu}^\dagger \phi V_{\mu} D^\dagger
V_{\mu}^\dagger \phi V_{\mu} DV_{\mu}^\dagger + $$
\begin {equation} 
                                                          \label {2.10}
D^\dagger V_{\mu}^\dagger \phi V_{\mu} D^\dagger  V_{\mu}^\dagger \phi -
V_{\mu}^\dagger \phi V_{\mu} D V_{\mu}^\dagger \phi V_{\mu}
D^\dagger V_{\mu}^\dagger =0
\end   {equation} 
Here $(\phi)_{ij}=\delta _{ij}\phi _{i}$. The solutions of this equation are
\begin {equation} 
                                                          \label {2.11}
V_{0}=\Upsilon P,
\end   {equation} 
where $\Upsilon$  is any diagonal unitary matrix and $P$ is any $N\times N$
permutation matrix which, when applied to an N-vector $\psi$, gives
\begin {equation} 
                                                          \label {2.12}
P_{ij}\psi _{j}=\psi _{P(i)}.
\end   {equation} 
Thus, for each permutation $P$, one has a solution of equation (\ref {2.10}).
So, for each $P$, we start with the decomposition of $V$ around $\Upsilon P$,
\begin {equation} 
                                                          \label {2.13}
V_{\mu}=V_{0\mu}{\cal U}_{\mu},
\end   {equation} 
where
\begin {equation} 
                                                          \label {2.14}
{\cal U}=1+i\lambda ^{\alpha}\xi _{\alpha}-\frac{1}{2} \lambda ^{\alpha}
\lambda ^{\beta}
\xi _{\alpha}\xi _{\beta} + {\cal O}(\xi ^{3})
\end   {equation} 
To get the complete answer in the semiclassical approximation, we have to
sum over all saddle points
\begin {equation} 
                                                          \label {2.15}
Z(\phi)=\prod _{\mu}\int dV_{\mu}e^{\tilde {S}(\Phi ,V{\mu})}
\sim \prod _{\mu}\sum _{P} \int _{around~P}d{\cal U}e^{a^{d}N\beta
S(\phi, D_{\mu},P,~{\cal U})}
\end   {equation} 
The effective action $S(\phi, D_{\mu},P,{\cal U})$ up to
quadratic terms on $\xi$ has the form
$$S(\phi _{i}, D_{\mu},P,{\cal U}) =\sum \phi _{P(i)}^{2} + $$
$$2\tr (\lambda \cdot \xi )\phi _{P} (\lambda \cdot \xi )\phi _{P}
+\tr (\lambda \cdot \xi )\phi _{P} D_{\mu}\phi _{P} (\lambda \cdot \xi )D_{\mu}
^\dagger $$
\begin {equation} 
                                                          \label {2.16}
+\tr (\lambda \cdot \xi )D_{\mu}(\lambda \cdot \xi )\phi _{P} D_{\mu}
^\dagger \phi _{P}
-2\tr (\lambda \cdot \xi )\phi _{P} D_{\mu}(\lambda \cdot \xi )\phi _{P}
D_{\mu}
^\dagger -2\tr (\lambda \cdot \xi )^{2}(\phi _{P}) ^{2}
\end   {equation} 
where $\phi _{P}$ is a diagonal matrix $diag\phi _{P}=\phi _{P(1)},\phi
_{P(2)},
...\phi _{P(n)}$; $\xi ^{\pm,ij}$ are the components of $(\lambda \cdot \xi )$
along of generators of U(N) (from this place for simplicity we deal with
$U(N)$)
\begin {equation} 
                                                          \label {2.18'}
(\lambda ^{+,ij})_{i'j'}=(\delta _{i'}^{i}\delta _{j'}^{j}-
\delta _{j'}^{i}\delta _{i'}^{j});~~
(\lambda ^{-,ij})_{i'j'}=i(\delta _{i'}^{i}\delta _{j'}^{j}+
\delta _{j'}^{i}\delta _{i'}^{j});~~
\end   {equation} 
(the components corresponding to generators being
diagonal matrices do not make a contribution in (\ref {2.17})).
Taking into account the formula
\begin {equation} 
                                                          \label {2.17}
\tr [A(\lambda \cdot \xi )B(\lambda \cdot \xi )-AB(\lambda \cdot \xi )
(\lambda \cdot\xi )]=
-\sum _{i<j}[A_{i}B_{j}+A_{j}B_{i}-A_{i}B_{i}-A_{j}B_{j}][(\xi ^{+,ij})^{2}+
(\xi ^{-,ij})^{2}],
\end   {equation} 
we get
\begin {equation} 
                                                          \label {2.18}
S(\phi _{i}, D_{\mu},P,{\cal U}) =\sum \phi _{P(i)}^{2} +
2\sum _{i<j}(\phi _{P(i)}-\phi _{P(j)})^{2}{\cal D}_{\mu ij}[(\xi ^{+,ij})^{2}+
(\xi ^{-,ij})^{2}]
\end   {equation} 
where
\begin {equation} 
                                                          \label {2.18a}
{\cal D}_{\mu ij}=(D_{\mu i}-D_{\mu j})(D ^\dagger _{\mu i}-D^\dagger _{\mu
j}).
\end   {equation} 
In the basis (\ref {2.18'}), the measure for   $U(N)$ takes the form
\begin {equation} 
                                                          \label {2.19a}
dV=\prod _{i<j} d\xi ^{+,ij} d\xi ^{-,ij} \exp [-\frac{1}{6}N
((\xi ^{+,ij})^{2}+(\xi ^{-,ij})^{2})]\prod _{i=1}^{N}d\xi _{i}
\exp [ -\frac{1}{6}N\xi _{i}^{2}+\frac{1}{6}(\sum _{i=1}^{N}\xi _{i})^{2}].
\end   {equation} 
Performing the gaussian integration over $\xi ^{\pm ,ij}, $
we get
\begin {equation} 
                                                          \label {2.19}
\prod _{\mu}\left(\sum _{P}e^{a^{d}N\beta \sum _{i}\phi _{P(i)}^{2}}
\prod _{i<j}\frac{1}{a^{d}N\beta(\phi _{P(i)}-\phi _{P(j)})^{2}
{\cal D}_{\mu ij} +\frac{N}{6}}\right).
\end   {equation} 
Having in mind that the distribution of the external momentum $p_{i}$ is
invariant under
permutation $i \to P(i)$ one can for any given $P$ renumerate the momentum and
therefore one can claim that all permutations give the same contribution,
so the final answer in semiclassical approximation up to some normalization
factor has a form
\begin {equation} 
                                                          \label {2.19'}
\exp (da^{d}N\beta\sum _{i}\phi _{i}^{2})
\prod _{\mu}\prod _{i<j}\frac{1}{a^{d}N
(\phi _{i}-\phi _{j})^{2}{\cal D}_{\mu ij}+\frac{N}{6\beta}}
\end   {equation} 
Including the contribution of the factor (\ref {2.19a}) in (\ref {2.19})
we mix the orders in the perturbation theory and the $O(g^{2})$ term
$(g^{2}=\beta ^{-1})$ is not correct. To get the correct $O(g^{2})$  answer the
two-loop corrections should be computed. Note that the expression (\ref
{2.19'})
is well defined for some equal eigenvalues of $\Phi$, that is in accordance
with
the well defined integral (\ref {2.8'}) over compact manifold, and only
singular point
for finite N is the point $\beta =\infty $, i.e. one cannot neglect
the contribution of the second term in the denominator of
(\ref {2.19'}). Of course, for infinite $N$ the
integral (\ref {2.8'}) can have a singular points (compare with phase
transitions
for the Gross-Witten  \cite {GW} and the Brezin-Gross  \cite {BG} models).
A hint to a phase transition  for $N=\infty$ gives a zero of the denominator.

Let us to compare the integral (\ref {2.8'}) with the Itzykson-Zuber integral
\cite {IZ}, where   one integrates the exponent of a
quadratic form on $V$. If we calculate the Itzykson-Zuber integral
 using the semiclassical approximation, then  the corresponding
 classical solutions have the same form, but
the contributions around given $P$ depend explicitly on $P$.
Summing over $P$ provides a
compensation of the singularities coming from coinciding eigenvalues of matrix
$\phi$, and in this  case there is no reason to mix the orders in
the perturbation theory, and, moreover, the remarkable fact is that summing
over
$P$ the leading terms of the semiclassical approximation gives the exact
expression
found by different methods  \cite {IZ,Metha}.

Substituting (\ref {2.19'}) in (\ref {2.3}) we get the vacuum energy as the
large $N$ limit of the following expression
\begin {equation} 
                                                          \label {2.20}
\frac{F}{(Vo~l)}=\int dp_{\mu}\ln [\int d\phi e^{Na^{d}(-m_{0}^{2}+\beta)
\sum \phi _{i}^{2}} \Delta ^{2}(\phi)\prod _{\mu}\prod _{i<j}
\frac{\sin ^{2}(\frac{p_{\mu}^{i}-p_{\mu}^{j}}{2}a)}{a^{d}N
(\phi _{i}-\phi _{j})^{2}{\cal D}_{\mu ij}+\frac{N}{6\beta}}].
\end   {equation} 

The free energy (\ref {2.20}) can be rewritten using the replica trick \cite
{PaR}. The
replica trick allows us to estimate
the free energy of the quenched system as the analytic continuation of the free
energy in a class of annealed systems with the action being the copy of n's
actions (\ref {2.20}). We have
$${\cal F}=\lim _{n\to 0}\frac{1}{n}
\left( \int dp\prod _{\alpha =1}
^{n} d\phi ^{\alpha} e^{Na^{d}(-m_{0}^{2}+\beta d)
\sum \phi _{i}^{\alpha 2}} \Delta ^{2}(\phi ^{\alpha})
\prod _{\mu}\prod _{i<j}
\frac{1}{a^{d}
(\phi _{i}^{\alpha}-\phi _{j}^{\alpha})^{2}{\cal D}_{\mu ij}
+\frac{1}{6\beta})} - 1\right)$$
\begin {equation} 
                                                      \label {2.22}
+const.
\end {equation} 

Equation (\ref {2.20}) points that there is a phase transition for some
$\beta =m_{0}^{2}\beta _{cr}$. The reason is the following. For  $\beta
=\infty$
 one has not the second term in the denominator of (\ref {2.20}) and we left
with the
answer which was expected in the naive continuous limit and which is unstable,
and for $\beta =0$ we get a stable model. So, it
is  natural to expect that the model exhibits a phase transition.

To estimate the order of $\beta _{cr}$ one can roughly estimate (\ref {2.22})
as
\begin {equation} 
                                                          \label {2.22'}
{\cal F}=\ln \int  d\phi _{i} e^{Na^{d}(-m_{0}^{2}+\beta d)
\sum \phi _{i}^{ 2}} \Delta ^{2}(\phi )
\prod _{\mu}\prod _{i<j}
\frac{1}{a^{d}\beta
<(\phi _{i}-\phi _{j})^{2}><{\cal D}_{\mu ij}>+\frac{1}{6}},
\end   {equation} 
where
$<(\phi _{i}-\phi _{j})^{2}>$  means  the average with semi-circular
distributions of eigenvalues
$\phi _{i}$  given by   $u(\phi )=\frac{m_{eff}}{2\pi}
\sqrt {4-m_{eff}^{2}\phi ^{2}}$, i.e. $$\frac{1}{N^{2}}\lim _{N\to \infty}
\sum _{i<j}
<(\phi _{i}-\phi _{j}^{2}> =$$
\begin {equation} 
                                                          \label {2.23}
\frac{1}{2}\int (\phi -\phi ')^{2}u(\phi)
u_{\alpha}(\phi ')d\phi d\phi '= \frac{1}{m_{eff}^{2}}=
\frac{1}{m_{eff}^{2}}.
\end   {equation} 
In our case $m_{eff}^{2}=a^{d}(m_{0}^{2}-d\beta)=a^{d}m_{0}^{2}(1-d\beta_{0})$,
$\beta _{0}$ is a dimensionless constant.

$<{\cal D}_{\mu ij}>$  means the average of (\ref {2.18a}) with
the measure (\ref {2.21}), so
$$<{\cal D}_{\mu ij}> =2(1-1/\sqrt{e}).$$

Therefore, the average of the first term in the  denominator is equal to
$$\frac{2(1-1/\sqrt{e})\beta _{0}}{1-d\beta _{0}}.$$ So, for $\beta _{0}=1/d$
the contribution of the first term becomes infinite and we can neglect the
second
term and we get unstable model what is in accordance with expected phase
transition
for $\beta _{0}=1/d$.

The region $\beta_{0}>1/d$ is forbidden, since
the action  becomes unbounded
from below. Let us examine the region $\beta_{0}<1/d$ including the negative
$\beta_{0}$. In this case the action is  bounded from below, but still
there is a reason for a phase transition due to zero of denominator in
(\ref {2.22'}) and
\begin {equation} 
                                                          \label {2.25}
\beta_{cr}=\frac{1}{d-4,8}.
\end   {equation} 
 This phase transition occurs only for $d < 4,8$.

 As for numerical factors we would like to note that they may be changed by
 a factor $1/J$ which takes into account the two-loop
corrections, i.e. in (\ref {2.19'}) must be $\frac{1}{J}$ instead of
$\frac{1}{6}$.
If $J$ is such that $0,8J>d$ then we can expect a phase transition. For
negative
$J$ the above estimation gives a phase transition without any restrictions on
d.

\section {Concluding Remarks}
In this paper an application of the quenched procedure to the KM model has been
considered. It has been noted that using the quenched procedure with
constraints
one obtains the one side model with the quartic interaction. This model has
been
investigated in the semiclassical approximation and the phase transition at
$\beta _{cr}=\frac{1}{d-4,8}$  has been found.

It will be shown in forthcoming paper \cite {IA92}  how it is possible to
modify the initial
model to get the reduced model with quenching which admits an analytical
treating
at the large $N$ limit using the technique of the orthogonal polynomials.

\end{document}